\documentclass{article}

\usepackage{arxiv}

\usepackage[utf8]{inputenc} 
\usepackage[T1]{fontenc}    
\usepackage{hyperref}       
\usepackage{url}            
\usepackage{booktabs}       
\usepackage{amsfonts}       
\usepackage{nicefrac}       
\usepackage{microtype}      
\usepackage{lipsum}
\usepackage{biblatex}
\usepackage{amsthm}
\usepackage{graphicx}
\usepackage{amsmath}
\usepackage{float}
\usepackage{multirow}
\usepackage{dcolumn}
\newcolumntype{d}[1]{D{.}{.}{#1}}
\usepackage{longtable}
\usepackage{threeparttable}
\graphicspath{ {./images/} }
\title{Overlapping Community Detection in Temporal Text Networks}

\newtheorem{definition}{Definition}
\author{
  Shuhan Yan, Yuting Jia, Xinbing Wang \\
  Shanghai Jiaotong University\\
  \texttt{\{799897466, hnxxjyt, xwang8\}@sjtu.edu.cn} \\
  
}

\begin{document}
\maketitle

\begin{abstract}
Analyzing the groups in the network based on same attributes, functions or connections between nodes is a way to understand network information. The task of discovering a series of node groups is called \textit{community detection}. Generally, two types of information can be utilized to fulfill this task, i.e., the link structures and the node attributes. The temporal text network is a special kind of network that contains both sources of information. Typical representatives include online blog networks, the World Wide Web (WWW) and academic citation networks. In this paper, we study the problem of overlapping community detection in temporal text network. By examining 32 large temporal text networks, we find a lot of edges connecting two nodes with no common community and discover that nodes in the same community share similar textual contents. This scenario cannot be quantitatively modeled by practically all existing community detection methods. Motivated by these empirical observations, we propose MAGIC (Model Affiliation Graph with Interacting Communities), a generative model which captures community interactions and considers the information from both link structures and node attributes. Our experiments on 3 types of datasets show that MAGIC achieves large improvements over 4 state-of-the-art methods in terms of 4 widely-used metrics.
\end{abstract}


\section{Introduction}
Network can serve as a powerful language to represent relational information among data objects from social, natural and academic domains.\parencite{yang2014detecting} Analyzing the groups in the network based on same attributes, functions or connections between nodes is a way to understand network information. Such groups of nodes can be users from the same organization in social networks\parencite{newman2004detecting}, proteins with similar functionality in bio- chemical networks\parencite{krogan2006global}, and papers from the same scientific fields in citation networks\parencite{ruan2013efficient}. The task of discovering a series of node groups is called \textit{community detection}. Traditional methods\parencite{newman2004detecting, schaeffer2007graph} mainly focus on finding disjoint communities and are all based on a restrictive assumption that each node can only belong to one single community. By relaxing this assumption, the overlapping community detection problem becomes more general and has attracted major attention recently\parencite{palla2005uncovering,xie2013overlapping}.

There are generally two types of information that can be utilized to discover overlapping communities\parencite{yang2013community}. The first one is the link structure, i.e., the presence and absence of edges. Classical methods\parencite{ahn2010link,airoldi2008mixed, palla2005uncovering} usually focus on this type of information and aim to extract a group of nodes with more links inside the group than between its members and outside the group\parencite{newman2006modularity}. The second type of information is the node attribute, including online profiles of users, pre-existing features of proteins, and textual contents of papers. Due to the prevalent noise in link structures, the approaches for detecting community based on both types of information \parencite{ruan2013efficient, yang2014detecting} have gained increasing popularity.

In this paper, we study the problem of overlapping community detection in temporal text networks. A temporal text network is a directed network in which each node has textual content and temporal information. Such networks are ubiquitous in the real world. Typical representatives include online blog networks, the World Wide Web (WWW), email correspondence networks, and academic citation networks. Identification of meaningful communities in temporal text networks provides useful knowledge for subsequent applications such as domain-specific ranking and user-targeted recommendation.

The contributions of our work are three-folded. First, we gather a collection of 32 temporal text networks with ground-truth communities. They enable us to derive insights of community structures and allow us to quantitatively evaluate community detection methods. Second, we study the interactions among ground-truth communities in temporal text networks and discover that a large proportion of nodes share a link due to community interactions. We also analyze how node attributes help to improve the quality of detected communities and find that nodes in the same communities share similar textual content. Third, based on the empirical observations, we propose MAGIC (Model Affiliation Graph with Interacting Communities), a probabilistic generative model which utilizes all sources of information in the temporal text network and scales to network with millions of nodes.

\paragraph{Present work: Networks with Ground-Truth Communities.}
We generate a large set of 32 temporal text networks with reliable ground-truth communities based on Microsoft Academic Graph (MAG)\footnote{We use the Feb 5, 2016 version of MAG.}\parencite{sinha2015overview}. The MAG dataset contains over 100 million scientific papers with titles, references, publish time, and sets of “Field of Study"(FoS). We construct a temporal text network by sampling an academic citation network for each L1 level FoS under Computer Science(CS) field and define each FoS label as a ground-truth community. Finally, we treat the publish time and title of each paper as its corresponding temporal and textual attributes. More
details about data preprocessing are presented in Section 4.1.

\paragraph{Present work: Empirical Observations.} The availability of temporal text networks with ground-truth communities enables us to derive insights of the community structure. In this paper, we study the interactions among ground-truth communities and discover that a large number of nodes share a link because of the community interactions. We find that a large proportion of edges connect a pair of nodes which have no communities in common. Current methods\parencite{yang2013overlapping, yang2013community, yang2014detecting}fail to identify such community interactions and fail to model this scenario well. We also quantitatively analyze how textual contents provide useful information for overlapping community detection. We find that nodes in the same communities share very similar textual content, which is very intuitive.

\paragraph{Present work: Community Detection in temporal text network.}
Based on the above empirical observations, we propose MAGIC (Model Affiliation Graph with Interacting Communities), a generative model which models the probability of an edge be- tween two nodes as a function of the communities they share, the interactions among communities they are affiliated in, and the time information of each node. MAGIC captures community interactions and considers the information from both link structures and node attributes. MAGIC further reduces the noise of missing links by utilizing the time information attached on each node. By fitting MAGIC toward a given temporal text network, we can detect meaningful communities. We conduct extensive experiments on 17 temporal text networks from three different sources–LFR benchmark\parencite{lancichinetti2009benchmarks}, SNAP, and MAG. Our results show that MAGIC achieves large improvements over 4 state-of-the-art methods \parencite{airoldi2008mixed,palla2005uncovering,yang2013overlapping,yang2013community} in terms of several widely-used metrics\parencite{lazar2010modularity, mcdaid2011normalized, yang2013overlapping}.

\paragraph{Organization}
The rest of this paper is organized as follows. Section 2 summarizes the related work. Section 3 gives formal definitions of temporal text network and the task of overlapping community detection. Section 4 presents our empirical observations. Section 5 introduces the MAGIC along with its learning method. Finally, we report the experimental results in section 6 and conclude in section 7.

\section{Related Work}

Overlapping community detection has been extensively investigated in the last decade\parencite{xie2013overlapping}. Classical methods such as CPM\parencite{palla2005uncovering}, MMSB\parencite{airoldi2008mixed}, and LC\parencite{ahn2010link} are mainly based on dense subgraph extraction. For example, CPM aims to find all k-cliques and combine those cliques sharing $k-1$ nodes to be communities. Consequently, these methods are not applicable for detecting communities in large-scale networks with millions of nodes.

More recently, a series of affiliation graph models \parencite{yang2012community, yang2013overlapping, yang2014structure, yang2014detecting} are proposed based on the idea that communities arise due to shared group affiliations\parencite{breiger1974duality}. Yang and Leskovec introduced Community-Affiliation Graph Model (AGM)\parencite{yang2012community} in which nodes are affiliated with latent communities they belong to and links are generated based on node community affiliations. They later relaxed the combinatorial optimization problem of fitting AGM and presented a more scalable model called BIGCLAM\parencite{yang2013overlapping}. This line of works, however, models the underlying affiliation network as a bipartite graph and assumes each community creates edges independently. Compared to these methods, MAGIC relaxes such assumption and captures community interactions.

Another piece of work which also considers community interactions is BNMTF\parencite{zhang2012overlapping}. This method factorizes the adjacency matrix of network into latent factors which are regarded as communities. However, BNMTF keeps using conventional Euclidean distance and generalized KL-divergence as the objective of matrix factorization, which is not scalable and causes bad interpretability.

Many models also study the problem of overlapping community detection in the context of combining link structure with node attributes\parencite{liu2015community, ruan2013efficient, yang2013community}. A large catalog of such models are based on topic models\parencite{balasubramanyan2011block, shen2016modeling, yang2009combining}. However, these methods do not allow a node to have high membership strength in multiple communities simultaneously and therefore leads to unrealistic assumptions about the structure of community overlaps. To solve this problem, authors in\parencite{yang2014structure} proposed CESNA which is an affiliation graph model based on BIGCLAM and uses a logistic model to generate binary-valued node attributes. CESNA models the generations of node attributes and link structures as two different mechanisms. Compared to CESNA, MAGIC takes a more unified approach to model these two types of information.

\section{Problem Formulation}
In this section, we formalize the problem of overlapping community detection in temporal text networks. We first define the “text network" and “temporal text network". Then, we discuss a method to explicitly encode text information in graph and define the “projected temporal text network". Figure 1 illustrates the relationship and difference among these three types of networks.

\begin{definition}[\textbf{Text Network}]
A \textbf{text network} is defined as a directed unweighted graph $G = (V, E)$, where $V$ is a set of vertices and E is the set of edges between the vertices. Each vertex $v \in V$ represents a document and has a sequence of words associated with it. Each edge $(u,v) \in E$ represents the directed connection between document $u$ and document $v$.
\end{definition}

The text network captures the relationship among documents and models it explicitly. Such network is ubiquitous in the real world. Online blog networks, email correspondence networks and academic citation networks are some good representatives.

\begin{definition}[\textbf{Temporal Text Network}]
A \textbf{temporal text network} is a text network with time information, denoted as $G = (V,E;T)$. In temporal text network, each vertex $v \in V$ is attached with a timestamp $t(v)$. Furthermore, a temporal text network is called natural temporal text network if each edge $(u,v)$ satisfies $t(u) < t(v)$; otherwise, it is called complex temporal text network.
\end{definition}
The temporal text network encodes the time information of each document. We state that most text networks are natural temporal text networks, provided that we give a proper definition of the edge direction. For example, if we define an edge in citation network starting from the cited paper to the citing one, then this network is natural because nobody can cite future papers.

\begin{definition}[\textbf{Projected Temporal Text Network}]
A \textbf{projected temporal text network}, denoted as $G^{p} = (V\cup V_w,E\cup E_{wd};T\cup T_w)$, is a transformation of original temporal text network $G = (V,E;T)$. Each additional vertex$v\in V_w$ represents a word and each additional edge $(w_i, d_j)\in E_{wd}$ indicates that word $w_i$ exists in document $d_j$. We set the timestamps of all word vertices to be zero\footnote{The exact number of this value is actually not important, as long as it is less than the earliest timestamp of all documents.}.
\end{definition}

Such projection method is proposed in\parencite{tang2015pte} and proved useful to model document-word dependency. The projected temporal text networks serves as a good proxy for the original network. Finally, we state that the problem investigated in this paper is overlapping community detection in temporal text networks. We will later elaborate their differences and discuss how we detect meaningful communities in the temporal text network by exploiting information in its corresponding projected version.

\begin{definition}[\textbf{Overlapping Community Detection in Temporal Text Networks}]
Given a temporal text network $G = (V,E;T)$, the problem of overlapping community detection in temporal text network is to find a collection of subsets of $V$ denoted by $C = {C_1,...,C_K}$ such that for each $C_i \in C$, its induced subgraph $G[C_i]$ forms a network community\footnote{An induced subgraph $G[C_i]$ is a graph whose vertex set is $C_i$ and whose edge set consists of all the edges in $E$ that have both endpoints in $C_i$ .}. By allowing $C_i\cap C_j \neq \emptyset$ we can obtain overlapping communities.
\end{definition}

\begin{figure}
  \centering
  \includegraphics[width=16cm]{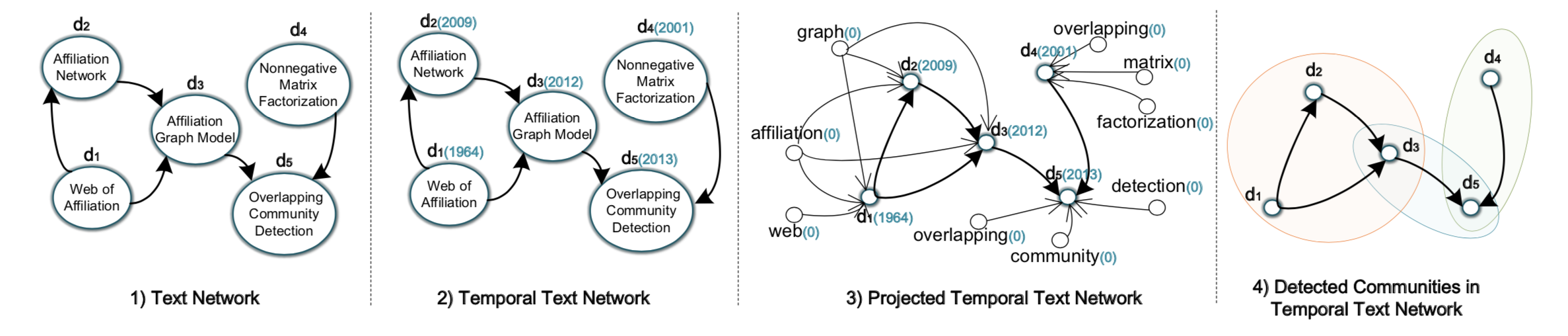}
  \caption{Illustration of three different networks. Blue-colored numbers in parentheses indicate the time information of each node. Clearly, there are two types of information in the temporal text network, i.e., the link structure and the node attribute. The projected temporal text network represents the text information in a more explicit manner and serve as a good proxy for its corresponding temporal text network.}
  \label{fig:pic1}
\end{figure}

\section{Empirical Observation}
In the section, we first describe how we generate a large collection of temporal text networks and define reliable ground-truth communities. Then, we present our empirical observations by answering two important questions. How many edges connect two nodes that share no common community? How textual contents improve the quality of detected communities? Finally, we discuss the importance of such findings and how they motivate the development of our model.

\subsection{Dataset descriptions}
We generate temporal text networks with explicit ground-truth communities based on Microsoft Academic Graph (MAG)\parencite{sinha2015overview}. The MAG dataset contains over 100 million scientific papers with their titles, references, publish time, and sets of “Field of Study"(FoS) labels. In total, there are over 50 thousands different FoS labels, organized in a four-level hierarchical manner as demonstrated in Figure 2a. Such FoS labels naturally correspond to ground-truth communities since all members (i.e., papers) of the same community are in the same subarea of science and possess the same property. Therefore, we define the FoS labels as the ground-truth communities and further treat the publish time and title of each paper as its temporal and textual attributes.

We construct a temporal text network by sampling an academic citation network. To illustrate the sampling process, we take the “Information Retrieval”(IR) field as an example. We consider that a paper is in IR field if it contains at least one FoS label in the set of IR-related FoS labels. A FoS is IR-related if it locates in the FoS tree rooted by the “Field of Study” named “Information Retrieval” (IR), as shown in Figure 2a. Then, we construct a citation network among all these selected papers and delete those with no reference and no citation. We repeat this process for 32 L1 level FoS under Computer Science (CS) field. These networks cover a wide range of domains and the sizes of them ranges from thousands of to millions of nodes.

\begin{figure}[H]
  \centering
  \includegraphics[width=16cm]{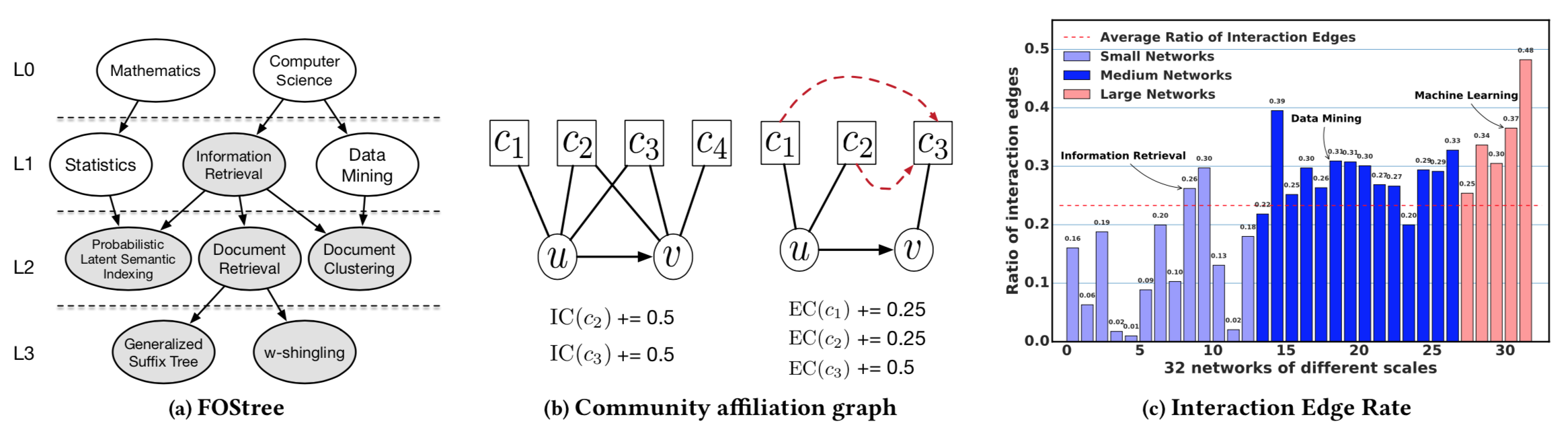}
  \caption{(a) A toy example of Field of Study (FoS) Tree. Each IR-related FoS is labeled with a shaded node. (b) An illustrative community affiliation graph. Circles represent nodes in the observed network. Squares represent the latent communities. IC denotes Internal Connectivity and EC denotes External Connectivity. (c) Ratio of interaction edges in networks with different scales.}
  \label{fig:pic2}
\end{figure}

\subsection{Empirical observation}
First, we analyze how textual contents help to provide useful information for community detection. For each community, we select two nodes and calculate the Jaccard similarity of their textual contents. The higher this value is, the more similar the textual contents are. We repeat this process for all possible pairs of nodes and get the average Jaccard similarity for that community. We compare this value with the average Jaccard similarity of a randomly selected set which has the same size of that community. Results are shown in Figure 3. As we can see, the average Jaccard similarity of each community is much higher than that of a randomly selected set.
This clearly demonstrates that nodes in the same community have
similar textual contents.

\begin{figure}[h]
  \centering
  \includegraphics[width=14cm]{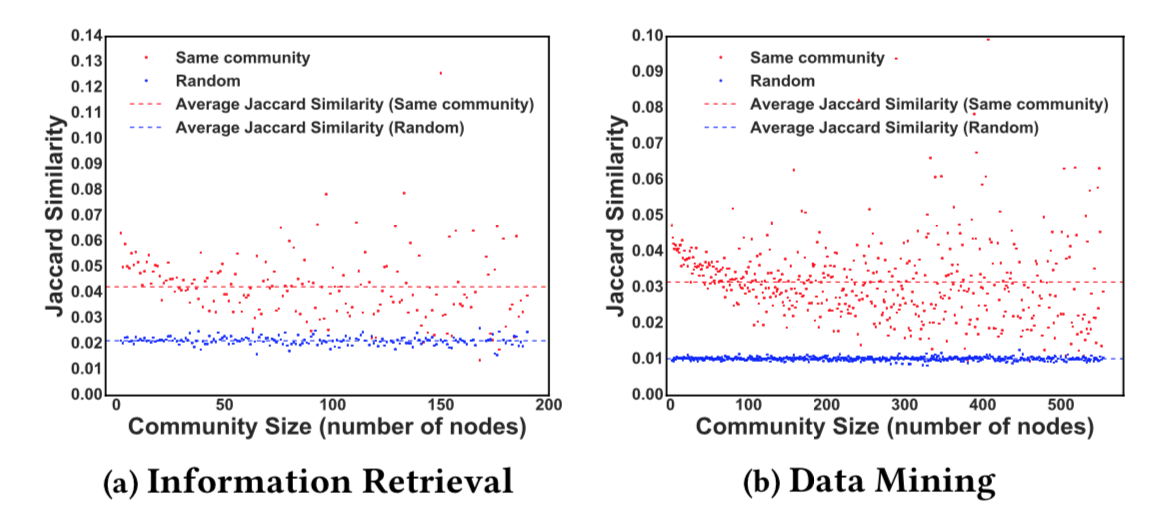}
  \caption{Comparison between the average Jaccard similarity of two randomly selected nodes and that of two nodes from the same community in two temporal text networks.}
  \label{fig:pic3}
\end{figure}

Next, we study community interactions by asking the question that how many edges connect two nodes that share no common community? These edges are caused by community interactions and thus we name them interaction edges. They reveal the overall amount of community interactions in each dataset. Results are shown in Figure 2c. As we can see, most of networks have more than 20\% of edges that are between two nodes with no common community. Besides, the ratio of such edge has an increasing trend with regard to the network size.

We then study such community interactions in a finer granularity. For each edge $(u,v)$, if node $u$ and $v$ have some communities in common, we assume this edge is generated only because two nodes share same communities. On the another hand, if node $u$ and $v$ have no common community, then this edge must be generated by community interactions. We formalize this idea as following. Let $C(u)$ denotes the set of communities of node u. For edge $(u,v)$, if $C(u)\cap C(v) \neq \emptyset$, then for each community $c\in C(u)\cap C(v)$, we add its Internal Connectivity (\textit{IC}) by $\frac{1}{\left | C(u)\cap C(v) \right |}$. Otherwise, if $C(u)\cap C(v) = \emptyset$, then for each community $c\in C(u)\cup C(v)$, we add its External Connectivity (\textit{EC}) by $\frac{1}{2 \times \left | C(u) \right |}$ if ($c\in C(u)$) and $\frac{1}{2 \times \left | C(v) \right |}$ if ($c\in C(v)$). An illustrative example is shown in Figure 2b.

After iterating all edges in the network, we can get IC score and
EC score of each community. We then define the \textit{Interaction Ratio}
of community c as $IR(c) = \frac{EC(c)}{IC(c)+EC(c)}$. Here we ignore all community interactions caused by edges linking two nodes that share some same communities.  Take Figure 2b as an example. We ignore possible community interactions between communities $c_1$ and $c_3$ or communities $c_2$ and $c_4$. We presume the edge $(u,v)$ is generated only due to the internal connectivity of community $c_2$ and $c_3$. Therefore, the Interaction Ratio measures the minimum amount of interaction for each community.

\begin{figure}[h]
  \centering
  \includegraphics[width=14cm]{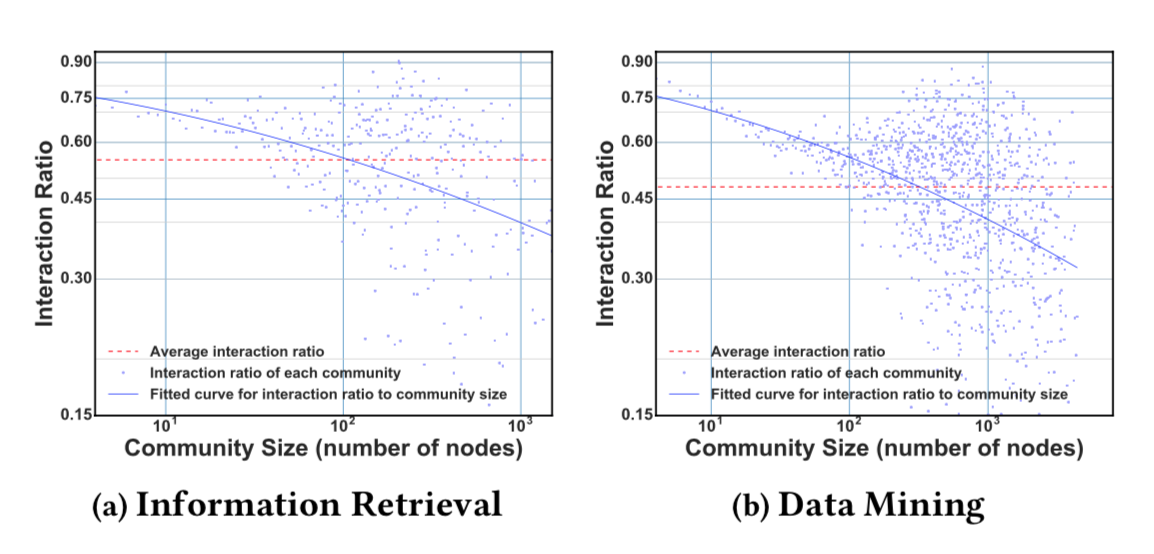}
  \caption{Interaction ratio of each community in four temporal text networks.}
  \label{fig:pic4}
\end{figure}

As shown in Figure 4, communities have strong interactions. This result is in retrospect, very intuitive. For example, papers in “Information Retrieval” field may adopt the techniques from “Natural Language Processing” papers for semantic search. An algorithm published in a “Machine Learning” conference has its origin from a “Mathematics” problem and been widely used in “Data Mining” field. Consequently, if we find two papers with one in “Data Mining” community and another in “Machine Learning” community, the probability that they share a link should not be modeled as zero, as practically all existing methods do \parencite{yang2014structure, yang2013community, yang2014detecting}. Instead, we should consider the community interactions and model them explicitly. 

\section{Community detection in temporal text network}
Motivated by previous observations, we present MAGIC (Model Affiliated Graph with Interacting Communities), a probabilistic generative model which models the community interactions explicitly.

\subsection{Model description}
MAGIC is based on the idea that communities arise due to shared group affiliation \parencite{breiger1974duality,fortunato2007resolution}, and views the whole network as a result generated by a variant of the community-affiliation graph model \parencite{yang2012community}. Same as the original one, MAGIC models the community affiliation strength between each pair of node $u$ and community $c$ with a nonnegative parameter $F_{uc}$ . MAGIC differs mainly in how we model the latent affiliation network. The original community-affiliation graph model treats the affiliation network as a bipartite graph, which fails to capture those important interactions among communities. MAGIC, instead, explicitly models the community interaction strength between every pair of community $c_i$ and $c_j$ with a nonnegative parameter $\eta_{ij}$ ($\eta_{ij} = 0$ indicates community $c_i$ and $c_j$ definitely have no relationship). Finally, we use the parameter $\eta_{ij}$ to model the probability that two nodes in the same community $c_i$ are connected.

Given those parameters, MAGIC generates a link $(u\rightarrow v)$ with the probability $p(u\rightarrow v)$ defined as follows:
\begin{equation}
    p(u \rightarrow v) = \sum_{i,j}{(1-\exp(F_{ui}\eta_{ij}F_{vj}))}\delta(u \rightarrow v) =1-\exp(-F_u^T\eta F_v)\delta(u \rightarrow v)
\end{equation}

where $F_u$ is a column vector representing the community affiliation strength for node $u$, $\eta$ is the community interaction matrix, and $\delta(u \rightarrow v)$ is the weighting function defined only on the timestamps of nodes $u$ and $v$. The introduction of $\eta$ and $\delta$ explicitly models the community interaction and utilizes the node temporal information. In this paper, we mainly focus on the natural temporal text network and thus the weighting function $\delta$ is defined as:

\begin{equation}
    \delta(u \rightarrow v) = \left( \begin{array}{rcl} 1 &if & t(u)<t(v) \\ 0 & otherwise \end{array}\right)
\end{equation}

Eq. 2 essentially restricts the generation of an edge starting from a node with early timestamp and ending with a node with later timestamp. This constraint is ubiquitous in the real world. We cannot cite a paper published in future nor forward an unreceived email.

Next, we discuss how MAGIC utilizes the text information. Instead of treating words and documents separately and use different mechanisms to generate them \parencite{yang2013community}, we adopt a more unified approach. We first construct a projected temporal text network corresponding to the original one and then applied MAGIC to this projected network. A projected temporal text network is intrinsically a heterogeneous network with two types of nodes – “document-node” and “word-node”. MAGIC treats them in the same way and will learn a feature vector representing the latent community affiliation strength for each document and word.

Finally, MAGIC learns the community affiliation matrix $F$ and the community interaction matrix $\eta$ by maximizing the log likelihood of the observed network G:

\begin{equation}
    \hat{F}, \hat{\eta}=\underset{F \geq 0, \eta \geq 0}{\operatorname{argmax}} \log P(G|F,\eta)
    =\underset{F \geq 0, \eta \geq 0}{\operatorname{argmax}} l(F, \eta)
\end{equation}
where nonnegative matrices $F\in \mathbb{R}^{K\times N}$, $\eta\in \mathbb{R}^{K\times K}$, $K,N$denote the number of communities and nodes, respectively. The log likelihood can be written out as below:

\begin{equation}
l(F, \eta)= =\sum_{(u \rightarrow v) \in E}{\log(1-\exp(-F_u^T\eta F_v))}-\sum_{\substack{(u \rightarrow v)\not\in E\\ t(u)<t(v)}}{F_u^T\eta F_v}
\end{equation}

Notice here we explicitly add the time constraint in the second term so that the absence of an edge $(u,v)$ with $t(u) \geq t(v)$ will not contribute to the likelihood. As demonstrated in Figure 5a, there are
two possible reasons for a missing link $(u,v)$. If we find $t(u)<t(v)$, which means this edge could have been generated, then the absence of such edge can provides some useful information and we define such edge as an unobserved link. Otherwise, if $t(u)\geq t(v)$, then the absence of this link carries no information because it cannot be generated anyway. Therefore, we define such edge as an impossible link. MAGIC only uses the information derived from observed links and unobserved links.

\begin{figure}[h]
  \centering
  \includegraphics[width=10cm]{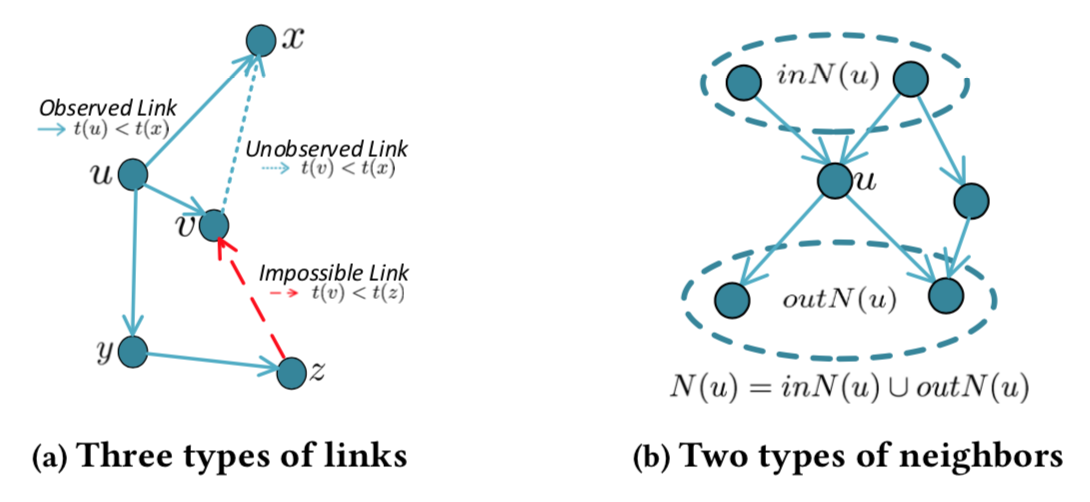}
  \caption{Interaction ratio of each community in four temporal text networks.}
  \label{fig:pic5}
\end{figure}

\subsection{Parameter learning}
We solve the optimization problem defined in Eq. 3 through block coordinate gradient ascent. We first update the community affiliation strength $F_u$ for each node $u$ with both $\eta$ and $F_v$ for all other nodes $v \neq u$ fixed, and then update the community interaction matrix $\eta$ with the community affiliation matrix $F$ fixed. To update the community affiliation strength $F_u$ for node $u$, we solve the following subproblem:

\begin{equation}
    \hat{F_u} = \underset{F_u\ge0}{\operatorname{argmax}} l(F_u)
\end{equation}

where $l(F_u)$ is the part of $l(F,\eta)$ defined in Eq. 4 that involves $F_u$, i.e.,

\begin{equation}
    l(F_u)=\sum_{v \in inN(u)}{\log(1-\exp(-F_v^T\eta F_u))}-\sum_{\substack{v\not\in N(u)\\ t(v)<t(u)}}{F_v^T\eta F_u}
    +\sum_{v' \in outN(u)}{\log(1-\exp(-F_u^T\eta F_v'))}-\sum_{\substack{v'\not\in N(u)\\ t(u)<t(v')}}{F_u^T\eta F_v'}
\end{equation}

where $inN(u)$ and $outN(u)$ denotes the set of in and out neighbors of node $u$, respectively. $N(u)$ is equal to $N(u)=inN(u)\cup outN(u)$, as demonstrated in Figure 5b. This subproblem can be further solved by projected gradient ascent\parencite{lin2007projected}.

\begin{equation}
    F_{uk}^{new} \leftarrow \max \{0, F_{uk}^{old}+\alpha_{F_u}(\nabla l(F_u))_k \}
\end{equation}

where $\alpha_{F_u}$ is the step size computed by backtracking line search\parencite{boyd2004convex}, and the gradient is:

\begin{equation}
\nabla l(F_u) = \sum_{v \in inN(u)}{\frac{\exp(-F_v^T\eta F_u)}{1-\exp(-F_v^T\eta F_u)} F_v^T\eta}-\sum_{\substack{v\not\in N(u)\\ t(v)<t(u)}}{F_v^T\eta}
    +\sum_{v' \in outN(u)}{\frac{\exp(-F_u^T\eta F_{v'})}{1-\exp(-F_u^T\eta F_{v'})}\eta F_{v'}} - \sum_{\substack{v'\not\in N(u)\\ t(u)<t(v')}}{\eta F_v'}
\end{equation}

After the parameter $F$ is updated, we fix $F$ and update the community interaction matrix $eta$. Notice that $\eta$ is involved in every term of Eq. 4 and thus we solve it directly.

\begin{equation}
    \eta_{ij}^{new} \leftarrow \max \{0, \eta_{ij}^{old}+\alpha_{\eta}(\nabla_{\eta} l(F,\eta))_{ij} \}
\end{equation}

where the step size $\alpha_{\eta}$ is also calculated by backtracking line search, and the gradient for $\eta$ is:

\begin{equation}
    \nabla_{\eta} l(F,\eta) = \sum_{(u \rightarrow v) \in E}{\frac{\exp(-F_u^T\eta F_v)}{1-\exp(-F_u^T\eta F_v)} F_u F_v^T} - \sum_{\substack{(u \rightarrow v)\not\in E\\ t(u)<t(v)}}{F_u F_v^T}
\end{equation}

We notice from Eqs.(8) and (10) that direct computations of $\nabla l(F_u)$ and $\nabla_{\eta} l(F,\eta)$ take $O(N)$ and $O(N^2)$ time,respectively. To reduce the time complexity and increase scalability, we adopt the following tricks:

\begin{equation}
    \sum_{v \notin N(u) \atop t(v)<t(u)} F_v^T \eta =\sum_{t(v)<t(u)} F_v^T \eta -\sum_{v \in i n N(u)} F_v^T \eta
\end{equation}
\begin{equation}
    \sum_{v' \notin N(u) \atop t(v')>t(u)} \eta F_{v'}=\sum_{t(v')>t(u)} \eta F_{v'}-\sum_{v' \in \text {out} N(u)} \eta F_{v'}
\end{equation}
\begin{equation}
    \sum_{(u \rightarrow v) \notin E} F_u F_v^T=\sum_{(u \rightarrow v)} F_u F_v^T-\sum_{(u \rightarrow v) \in E} F_u F_v^T
\end{equation}

In this way, we can compute $\nabla l(F_u)$ in $O(N(u))$ by caching the first term in the right hand side of Eqs.(11) and (12), and compute $\nabla_{\eta} l(F,\eta)$ in $O(|E|)$ by caching the first term in the right hand side of Eq.13. We notice that the combined time complexity for updating the whole $F$ is $O(|E|)$. Therefore, we conclude that the time complexity of each iteration for MAGIC is $O(|E|)$.

\subsection{Other issues}

\paragraph{Model initialization.} To initialize $F$, we extend the method in \parencite{gleich2012vertex} to directed network. The conductance in directed network in also defined in \parencite{galhotra2015tracking}. The in-neighbors $inN(u)$ of node $u$ is locally minimal if $inN(u)$ has lower conductance than all in-neighbors $inN(v)$ where node $v \in outN(u)$. For a node $u'$ belonging to such a locally minimal neighborhood $k$, we initialize $F_{u'k} =1$, otherwise we let $F_{u'k} = 0$. To initialize $\eta$, we set the entries in the main diagonal as 0.9 and all other entries to be 0.1.

\paragraph{Determining community membership.} After learning parameters $\hat{F}$ and $\hat{\eta}$, we need to determine the “hard” community membership of each node. We achieve this by thresholding $\hat{F}$ with a set of $\{\delta_{k}\}$, one for each community $c_k$. The basic intuition is that if two nodes belong to the same community $c_k$, then the probability of having an link between them through community $c_k$ is larger than $1/N$ , where $N$ is the number of nodes. Following this idea, we can obtain $\delta_{k}$ as below:

\begin{equation}
\delta_{k}=\sqrt{-\frac{\log (1-1/N)}{\eta_{k k}}}
\end{equation}

With $\{\delta_{k}\}$ obtained, we consider node $u$ belonging to community $k$ if $F_{uk} \ge \delta_k$.

\paragraph{Choosing the number of communities.} We use the method in \parencite{airoldi2008mixed} to choose the number of communities $K$. Specifically, we reserve 20\% of links for validation and learn the model parameters with the remaining 80\% of links for different $K$. After that, we use the learned parameters to predict the links in validation set and select the $K$ with the maximum prediction score as the number of communities.

\section{Experiments}
\subsection{Experiment setup}
\subsubsection{Dataset}
We evaluate our model using three categories of networks with ground-truth communities. In all networks, each node is assigned to at least one community and thus all quantitative metrics are applicable.

\textbf{MAG (Microsoft Academic Graph)}
We constructed 32 temporal text network based on the MAG as described in the Section 4, of which we randomly choose 11 for the experiment. The source codes and cleaned datasets are available online\footnote{http://tinyurl.com/WSDM2017-205}.

\textbf{SNAP (Standard Network Analysis Project)} 
SNAP provides six networks with ground-truth communities, among which we choose three, the Amazon product co-purchasing network, the Youtube social network and the LiveJournal blogging community\footnote{http://snap.stanford.edu}. For the experiment, we create 500 different subnetworks for each of the two datasets as in \parencite{yang2013overlapping}.

\textbf{LFR (Lancichinetti-Fortunato-Radicchi)} 
Benchmark LFR \parencite{lancichinetti2009benchmarks} is one of the state-of-the-art synthetic networks with overlapping ground-truth communities. In our experiments, we change parameters N-D-R to generate three different networks, named LFR Small/Middle/Large. Here N is the number of nodes in the network, D is the average degree and R is the ratio of nodes which belong to multiple communities. For the LFR Small/Middle/Large, we set N-D-R as (10k-8-0.4), (50k-12-0.6), (100k-16-0.8) respectively.

\subsubsection{Comparative Methods}
We compare MAGIC with another 4 baseline methods: Clique Percolation Method (CPM) \parencite{palla2005uncovering}, Mixed- Membership Stochastic Block Model (MMSB)\parencite{airoldi2008mixed}, Cluster Affiliation Model for Big Networks (BIGCLAM)\parencite{yang2013overlapping}, and Communities from Edge Structure and Node Attributes (CESNA)\parencite{yang2013community}. CPM and MMSB are representatives of methods based on dense subgraph extraction while BIGCLAM and CSENA are representatives of methods based on affiliation graph model. For CPM, we set the clique size $k = 4$ and use the implementation in the Stanford Network Analysis Platform (SNAP)\footnote{https://github.com/snap-stanford/snap}. In addition to CPM, SNAP also provides implemention of BIGCLAM anf CESNA and we adopt them for the experiment. For MMSB, We use the implementation in \parencite{gopalan2013efficient}\footnote{MMSB: https://github.com/premgopalan/svinet}.

Previously BIGCLAM has been shown to outperform NMF \parencite{psorakis2011overlapping,wang2011community} and CSENA has been shown to outperform CODICIL \parencite{ruan2013efficient}  and Block-LDA\parencite{balasubramanyan2011block}[21]. Therefore, we do not compare with those algorithms.

\subsubsection{Metrics}
We denote the set of ground-truth communities as $C$ and the set of detected communities as $\hat{C}$. To measure the performance of our model, we select following 4 metrics.

\textbf{Coverage ratio} is the ratio of nodes which can be assigned to at least one community by the model. Intuitively, a model cannot be useful if it can only detects communities for a very few proportion of nodes.

\textbf{F1 score} is the average of the F1 score of the best-matching ground-truth community to reach detected community. Please refer to \parencite{yang2013overlapping} for details.

\textbf{Modularity} is an unsupervised metric for communities. We use the extended version of the modularity to apply it on the overlapping communities. Please refer to \parencite{lazar2010modularity} for details.

\textbf{Omega index} is a metric with the basic idea of estimating the number of communities that each pair of nodes shares. Please refer to \parencite{yang2013overlapping} for details.

For all 4 metrics higher values mean that the detected communities are more accurate and have better qualities.

\subsection{Quantitative results}
\subsubsection{Performance on three categories of network}

Table 1 shows the performance of 5 methods on 9 networks in terms of 4 metrics. These 9 networks belong to 3 different types. And only MAG net- works contain textual content so that for CESNA we only measure it on MAG. What’s more, SNAP networks are undirected network so that for model MAGIC we use MAGIC(all) for MAG, MAGIC(net) for LFR and MAGIC(raw) for SNAP and all of these versions are referred to as MAGIC in the table.

\begin{table}[!htb]
\caption{Results of 5 methods on 9 networks from 3 different categories.}
  \label{test1}
 \centering
\resizebox{\textwidth}{!}{ %
\begin{tabular}{clccccccccr}
\multirow{2}{*}{\textbf{Metric}}                                           & \multicolumn{1}{c}{\multirow{2}{*}{\textbf{Method}}} & \multicolumn{3}{c}{\textbf{MAG}}                                                                                                                                                                                                                                                     & \multicolumn{3}{c}{\textbf{LFR}}                                                                              & \multicolumn{3}{c}{\textbf{SNAP}}                                                                                                                                 \\ \cline{3-11} 
                                                                           & \multicolumn{1}{c}{}                                 & \multicolumn{1}{l}{\textbf{\begin{tabular}[c]{@{}l@{}}Internet \\ Privacy\end{tabular}}} & \multicolumn{1}{l}{\textbf{\begin{tabular}[c]{@{}l@{}}Computer \\ Hardware\end{tabular}}} & \multicolumn{1}{l}{\textbf{\begin{tabular}[c]{@{}l@{}}Information\\  Retreival\end{tabular}}} & \multicolumn{1}{l}{\textbf{Small}} & \multicolumn{1}{l}{\textbf{Middle}} & \multicolumn{1}{l}{\textbf{Large}} & \multicolumn{1}{l}{\textbf{Amazon}} & \multicolumn{1}{l}{\textbf{Youtube}} & \multicolumn{1}{l}{\textbf{\begin{tabular}[c]{@{}l@{}}Live \\ Journal\end{tabular}}} \\ \hline
\multirow{5}{*}{\begin{tabular}[c]{@{}c@{}}Coverage \\ ratio\end{tabular}} & MAGIC                                                & \textbf{0.998*}                                                                          & \textbf{1.000*}                                                                           & \textbf{1.000*}                                                                               & \textbf{0.999*}                     & \textbf{1.000*}                      & \textbf{0.999*}                     & \textbf{0.996*}                      & 0.832                                & 0.972                                                                                \\
                                                                           & CPM                                                  & 0.271                                                                                    & 0.017                                                                                     & 0.047                                                                                         & 0.053                              & 0.019                               & 0.008                              & 0.813                               & 0.233                                & 0.939                                                                                \\
                                                                           & MMSB                                                 & 0.823                                                                                    & 0.799                                                                                     & 0.799                                                                                         & 0.908                              & 0.967                               & 0.984                              & 0.985                               & \textbf{0.836*}                      & \textbf{0.983*}                                                                      \\
                                                                           & BIGCLAM                                              & 0.861                                                                                    & 0.561                                                                                     & 0.38                                                                                          & 0.611                              & 0.409                               & 0.307                              & 0.982                               & 0.671                                & 0.953                                                                                \\
                                                                           & CESNA                                                & 0.875                                                                                    & 0.716                                                                                     & 0.838                                                                                         & -                                  & -                                   & -                                  & -                                   & -                                    & -                                                                                    \\ \hline
\multirow{5}{*}{\begin{tabular}[c]{@{}c@{}}F1 \\ score\end{tabular}}       & MAGIC                                                & \textbf{0.235*}                                                                          & \textbf{0.175*}                                                                           & \textbf{0.171*}                                                                               & \textbf{0.245*}                    & \textbf{0.061*}                     & \textbf{0.019*}                    & \textbf{0.173*}                     & \textbf{0.070*}                      & 0.063                                                                                \\
                                                                           & CPM                                                  & 0.137                                                                                    & 0.008                                                                                     & 0.021                                                                                         & 0.050                              & 0.008                               & 0.005                              & 0.094                               & 0.011                                & 0.056                                                                                \\
                                                                           & MMSB                                                 & 0.199                                                                                    & 0.053                                                                                     & 0.087                                                                                         & 0.147                              & 0.03                                & 0.011                              & 0.153                               & 0.046                                & 0.060                                                                                \\
                                                                           & BIGCLAM                                              & 0.208                                                                                    & 0.104                                                                                     & 0.073                                                                                         & 0.157                              & 0.036                               & 0.019                              & 0.168                               & 0.055                                & \textbf{0.071*}                                                                      \\
                                                                           & CESNA                                                & 0.210                                                                                    & 0.107                                                                                     & 0.121                                                                                         & -                                  & -                                   & -                                  & -                                   & -                                    & -                                                                                    \\ \hline
\multirow{5}{*}{Modularity}                                                & MAGIC                                                & 0.506                                                                                    & 0.500                                                                                     & 0.498                                                                                         & \textbf{0.500*}                    & \textbf{0.500*}                     & \textbf{0.500*}                    & \textbf{0.614*}                     & \textbf{0.543*}                      & 0.665                                                                                \\
                                                                           & CPM                                                  & 0.411                                                                                    & 0.436                                                                                     & 0.418                                                                                         & 0.337                              & 0.349                               & 0.263                              & 0.607                               & 0.233                                & \textbf{0.845*}                                                                      \\
                                                                           & MMSB                                                 & 0.518                                                                                    & 0.501                                                                                     & 0.500                                                                                         & 0.500                              & 0.500                               & 0.500                              & 0.581                               & 0.518                                & 0.682                                                                                \\
                                                                           & BIGCLAM                                              & 0.538                                                                                    & \textbf{0.561*}                                                                           & \textbf{0.550*}                                                                               & 0.500                              & 0.500                               & 0.500                              & 0.570                               & 0.522                                & 0.642                                                                                \\
                                                                           & CESNA                                                & \textbf{0.542*}                                                                          & 0.560                                                                                     & 0.546                                                                                         & -                                  & -                                   & -                                  & -                                   & -                                    & -                                                                                    \\ \hline
\multirow{5}{*}{Omega}                                                     & MAGIC                                                & \textbf{0.426*}                                                                          & \textbf{0.406*}                                                                           & 0.508                                                                                         & 0.886                              & 0.797                               & 0.683                              & 0.389                               & \textbf{0.422*}                      & 0.120                                                                                \\
                                                                           & CPM                                                  & 0.388                                                                                    & 0.366                                                                                     & 0.559                                                                                         & \textbf{0.969*}                    & \textbf{0.987*}                     & \textbf{0.989*}                    & 0.277                               & 0.132                                & 0.008                                                                                \\
                                                                           & MMSB                                                 & 0.392                                                                                    & 0.350                                                                                     & 0.561                                                                                         & 0.896                              & 0.819                               & 0.750                              & \textbf{0.401*}                     & 0.403                                & 0.163                                                                                \\
                                                                           & BIGCLAM                                              & 0.395                                                                                    & 0.347                                                                                     & 0.562                                                                                         & 0.946                              & 0.965                               & 0.971                              & 0.460                               & 0.375                                & \textbf{0.201*}                                                                      \\
                                                                           & CESNA                                                & 0.397                                                                                    & 0.350                                                                                     & \textbf{0.564*}                                                                               & -                                  & -                                   & -                                  & -                                   & -                                    & -                                                                                    \\ \hline
\end{tabular}}
\end{table}
 
 First, we notice that the coverage ratio of MAGIC in most networks is greater than 0.996, which means it can label almost all nodes with at least one community. In addition, the F1 score of MAGIC is the highest compared to other methods except in one network. The remaining two metrics are bias to models without considering community interaction for the reason that they only focus on edges or nodes in the same community. But for our model, even though two nodes belong to different communities, it’s also possible for them to be linked because of community interaction. Due to this property of the two metrics, MAGIC doesn’t get the highest score.
 
\subsubsection{Scaled performance comparison}
According to the results in Table 1, we can see different metrics measure different aspects community detection algorithms and it’s hard to find a method that totally outperform others for all metrics. To compare the overall performance of a method, we need to normalize those metrics to the same scale. Specifically, we scale it to make sure the best community detection method will get a score 1. Then, we sum up all four normalized scores to obtain the composite score. In order to run the integrated MAGIC, i.e., MAGIC(all), we only randomly choose 7 networks from MAG dataset which contain temporal and textual information.

Figure 6 compares the performance of 6 methods on 7 MAG networks in terms of the sum of four normalized metrics. We can see that MAGIC(all) achieves the best performance in all the 7 networks and mostly MAGIC(net) gets the second place. The average composite performance of MAGIC(all) is 3.68, which is 41\% higher than BIGCLAM(2.61), 19\% higher than CESNA(3.11), 138\% higher than CPM(1.55), and 33\% higher than MMSB(2.88).

\begin{figure}[h]
  \centering
  \includegraphics[width=16cm]{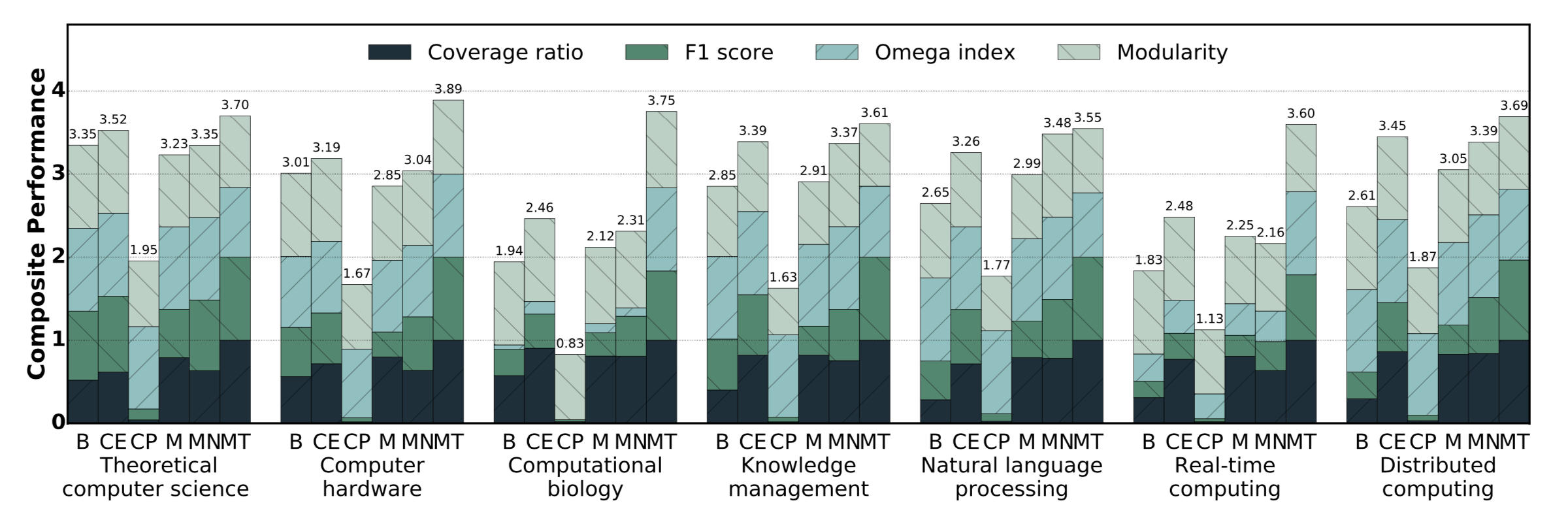}
  \caption{The composite performance of 6 methods on 7 small scale temporal text networks. B: BIGCLAM; CE: CESNA, CP: CPM, M: MMSB, MN: MAGIC(net), MA: MAGIC(all).}
  \label{fig:pic6}
\end{figure}

\subsection{Effects of community interactions}
We further analyze how community interactions affect quality of detected communities when different combinations of information sources are used. To achieve this, we introduce a new method called CoDA (Communities through Directed Affiliations)\parencite{yang2014detecting}, an overlapping community detection method that applies to directed networks. Totally, we have six methods to compare and we divide them into three groups. Table 2 shows the composite performance of these methods. Since the Amazon network and the Youtube network are undirected and without node attributes, only BIGCLAM and MAGIC(raw) are able to run on them. Likewise, CESNA and MAGIC(all) can not process the LFR networks due to the absence of node attributes in the synthetic networks.

\begin{table}[!htb]
\caption{Results of six methods on 7 networks of 3 different types.}
  \label{test2}
 \centering
 \resizebox{\textwidth}{!}{ %
\begin{tabular}{lcccccccccc}
\multicolumn{1}{c}{\multirow{2}{*}{\textbf{Method}}} & \multirow{2}{*}{\textbf{Directed}} & \multirow{2}{*}{\textbf{Text}} & \multirow{2}{*}{\textbf{\begin{tabular}[c]{@{}c@{}}Community \\ Interaction\end{tabular}}} & \multicolumn{3}{c}{\textbf{MAG}}                                                                                                                                                                                                                 & \multicolumn{2}{c}{\textbf{LFR}}                                        & \multicolumn{2}{c}{\textbf{SNAP}}                                          \\ \cline{5-11} 
\multicolumn{1}{c}{}                                 &                                    &                                &                                                                                            & \textbf{\begin{tabular}[c]{@{}c@{}}Computer \\ Graphics\end{tabular}} & \textbf{\begin{tabular}[c]{@{}c@{}}Knowledge \\ Management\end{tabular}} & \multicolumn{1}{l}{\textbf{\begin{tabular}[c]{@{}l@{}}Information \\ Retreival\end{tabular}}} & \multicolumn{1}{l}{\textbf{Small}} & \multicolumn{1}{l}{\textbf{Large}} & \multicolumn{1}{l}{\textbf{Amazon}} & \multicolumn{1}{l}{\textbf{Youtube}} \\ \hline
BIGCLAM                                              & F                                  & F                              & F                                                                                          & 3.042                                                                 & 2.853                                                                    & 2.702                                                                                         & 3.414                              & 3.291                              & \textbf{3.885*}                     & 3.43                                 \\
MAGIC(raw)                                           & F                                  & F                              & T                                                                                          & \textbf{3.117*}                                                       & \textbf{3.109*}                                                          & \textbf{3.219*}                                                                               & \textbf{3.958*}                    & \textbf{3.720*}                    & 3.846                               & \textbf{3.995*}                      \\ \hline
CoDA                                                 & T                                  & F                              & F                                                                                          & 3.145                                                                 & 3.128                                                                    & 3.246                                                                                         & 3.499                              & 3.212                              & -                                   & -                                    \\
MAGIC(net)                                           & T                                  & F                              & T                                                                                          & \textbf{3.463*}                                                       & \textbf{3.607*}                                                          & \textbf{3.351*}                                                                               & \textbf{3.913*}                    & \textbf{3.690*}                    & -                                   & -                                    \\ \hline
CESNA                                                & F                                  & T                              & F                                                                                          & 3.41                                                                  & 3.389                                                                    & 3.377                                                                                         & -                                  & -                                  & -                                   & -                                    \\
MAGIC(all)                                           & T                                  & T                              & T                                                                                          & \textbf{3.533*}                                                       & \textbf{3.607*}                                                          & \textbf{3.574*}                                                                               & -                                  & -                                  & -                                   & \multicolumn{1}{l}{-}                \\ \hline
\end{tabular}}
\end{table}

In the first group, we ignore all node attributes (both temporal information and textual contents) as well as edge directions, and compare the results of BIGCLAM and MAGIC(raw). We can see MAGIC(raw) beats BIGCLAM in six out of seven networks. The average performance of MAGIC(raw) is 3.57, which is about 10\% higher that of BIGCLAM (3.23). Such improvements occur in another two groups where we use the edge directions and text information, respectively. We contribute these improvements to the introduction of community interactions as the only variable in all three groups is whether such interactions are considered or not.

\subsection{Word community}

Using MAGIC, we can actually obtain both document and word communities for each temporal text network. Those word communities have semantic meanings. Here, we present four example word communities from the “Theoretical Computer Science” field. We choose the field “Theoretical computer science” as it covers a broad range of subjects including “Computational biology", “Computer network", and etc. Meaningful word communities can be observed in Figure 7.

\begin{figure}[h]
  \centering
  \includegraphics[width=10cm]{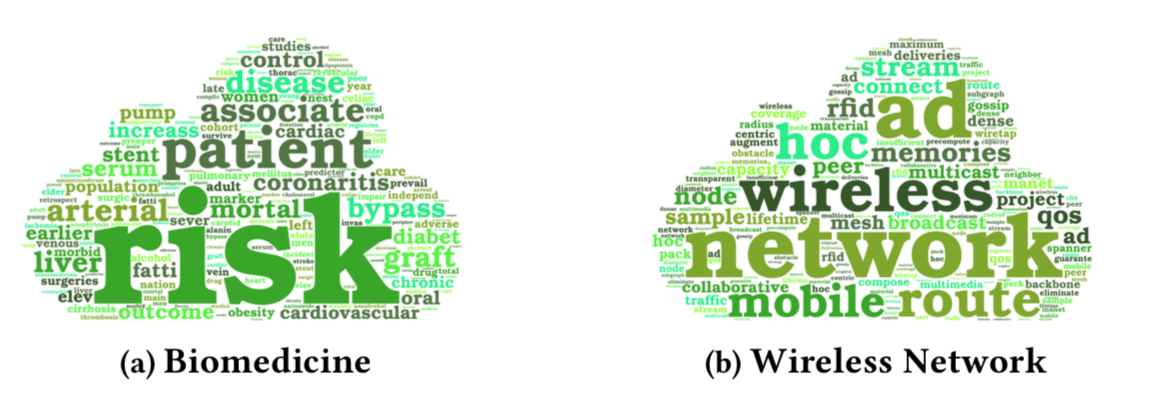}
  \caption{Word Clouds of Word Communities}
  \label{fig:pic7}
\end{figure}

\section{Conclusion}
In this paper, we study the problem of overlapping community detection in temporal text networks. We generate a large set of 32 temporal text networks with reliable ground-truth communities. They enable us to quantitatively study the community structure and evaluate community detection methods. We study the interactions among communities and discover that many nodes share a link due to such community interactions. We also find that nodes in the same community have similar textual contents. Based on these empirical observations, we propose MAGIC, a generative model which explicitly models the community interactions and utilizes the information from both link structures and node attributes. Extensive experiments based on 17 networks from 3 different sources demonstrate the effectiveness of MAGIC.

\printbibliography
\nocite{*}

\end{document}